\documentclass[onecolumn,showpacs,preprintnumbers]{revtex4}
\usepackage{epsfig,graphicx,times}
\usepackage{amstext}
\usepackage{amsmath}
\usepackage{amssymb}
\usepackage{graphicx}
\usepackage{latexsym}
\usepackage{bm}
\usepackage{color}
\usepackage{epstopdf}

\begin{document}

\title{ Testing quantum adiabaticity with quench echo}

\author{H. T. Quan and W. H. Zurek}
\email{whzurek@gmail.com} \homepage{http://public.lanl.gov/whz/}
\affiliation{Theoretical Division, MS B213, Los Alamos National
Laboratory, Los Alamos, NM, 87545, U.S.A.}

\begin{abstract}
Adiabaticity of quantum evolution is important in many settings; one
example is the adiabatic quantum computation (AQC). Nevertheless, to
date, there is no effective method available for testing the
adiabaticity of the evolution for the case where the eigenenergies
of the driven Hamiltonian are not known. We propose a simple method
for checking adiabaticity of a quantum process for an arbitrary
quantum system. We further propose an operational method for finding
more efficient protocols that approximate adiabaticity, and suggest
a ``uniformly adiabatic'' quench scheme based on the Kibble-Zurek
mechanism for the case where the initial and the final Hamiltonians
are given. This method should help in implementing AQC and other
tasks where preserving the system in the ground state of a
time-dependent Hamiltonian is desired.
\end{abstract}

\maketitle

\section{Introduction}

In a quantum quench process, when the Hamiltonian of a quantum
system is driven from $H_{0}$ to $H_{1}$, interstate excitations of
the system usually occur, owing to the non-commutativity of the
Hamiltonians at different moments. However, when the quench process
is slow enough, the interstate excitations will be suppressed.
According to the quantum adiabatic theorem \cite{adiabatictheorem},
when the condition for quantum adiabatic approximation $\left\langle
\Phi_{\mathrm{ground}} (t)| \frac{d H(t)}{dt} |
\Phi_{\mathrm{excited}}(t)\right \rangle \ll \Delta(t)^{2}$ is
satisfied, the system will remain in the ground state -- its
evolution will be adiabatic -- except for some special situations
\cite{clarification}. Here $H(t)$ is the changing Hamiltonian, and
$\Delta(t)$ is the minimal energy gap between the ground state
$\left | \Phi_{\mathrm{ground}} (t)\right\rangle$ and the first
excited state $\left | \Phi_{\mathrm{excited}} (t)\right\rangle$ of
$H(t)$.

In order to ensure that a quantum system evolves adiabatically, one
usually needs to find the energy spectrum (or at least the smallest
energy gap $\Delta$) of the driven Hamiltonian. One can then use the
quantum adiabatic theorem to choose a proper time scale, so that the
conditions for quantum adiabatic approximation are satisfied and the
evolution remains adiabatic. Nevertheless, in practice, neither
eigenenergies nor eigenstates of a complex quantum many-body system
are easy to obtain. This is often the case in implementing the
adiabatic quantum computation (AQC) \cite{qac} as well as quantum
annealing \cite{nishimori98,anarb2008}. Hence, one does not have the
ingredients to use quantum adiabatic theorem. One cannot count on
the direct comparison between the final state and the instantaneous
ground state of the final Hamiltonian, either.
Thus, it would be useful to find a reliable method for evaluating
the adiabaticity of an evolution under an arbitrary Hamiltonian,
especially when one has no idea about the eigenstates and/or
eigenenergies of the system (except at the initial moment).

The quench echo method we propose here is one solution to the above
problem. It will allow one to evaluate unambiguously the
adiabaticity of a process. What is more, it can help one find the
efficient ``uniformly adiabatic'' quench path in the parameter space
of the Hamiltonian. Such ideas may have applications in the
implementation of AQC \cite{qac}. This paper is organized as
follows: In section II, we introduce the quench echo method and
briefly explain its underlying physics. In section III, we use a
simple model to demonstrate main ideas of the general theory. In
Section IV, we propose a uniformly adiabatic scheme that is based on
the application of the Kibble-Zurek mechanism (KZM) to quantum phase
transitions. In Section V we give discussions and conclusions.

 \section{Quench echo}
Consider a system described by the Hamiltonian $H(g(t))$, where
$g(t)$ is a time-dependent parameter. The system is initially
prepared in the ground state of $H(g(t=0))$. The system evolves
under the influence of the driven Hamiltonian, which changes from
$H(g(t=0))$ to $H(g(t=T))$. Our aim is to test the adiabaticity of
this evolution, but we know nothing about the eigenenergies and
eigenstates of the time-dependent Hamiltonian except at $t=0$. Hence
we cannot count on the comparison between the evolving state and the
instantaneous ground state. Neither can we use the adiabatic
theorem. Nevertheless, we can apply a backward ``echo" quench
following the initial quench (from $t=0$ to $t=T$). That is, one
extends the evolution from $t=T$ to  $t=2T$ \cite{footnote}:
\begin{equation}
H(g(t))= \left \{
\begin{array}{c}
\begin{split}
&H(g(t)), (0<t<T)  \\
&H(g(2 T-t)), (T <t< 2 T)
\end{split}
\end{array}
\right. .\label{1}
\end{equation}
The final Hamiltonian is identical to the initial Hamiltonian
$H(g(t=2 T)) \equiv H(g(t=0))$. Hence, we can use the fidelity of
the initial state, e.g., the ground state of $H(g(t=0))$ and the
final evolving state as a criterion for the adiabaticity of the
evolution.
\begin{equation}
F=\left | \left \langle GS \right | \hat {T} e^{-i
\int_{T}^{2T} H(g(2 T -t)) dt} \hat {T} e^{-i
\int_{0}^{T} H(g(t)) dt} \left | GS \right \rangle \right |^{2},
\label{2}
\end{equation}
where $\hat {T}$ is the time-ordered operator and $\left | GS \right
\rangle$ is the ground state of $H(g(t=0))$. When the fidelity $F$
is greater than a threshold value close to unity, (e.g., $0.999$,
the error tolerance is $0.001$), the whole process $(0< t < 2 T)$ is
adiabatic. This implies that the forward quench process $(0< t < T)$
is adiabatic. The underlying mechanism for this ``quench echo"
method is straightforward: Except for the eigenstate of the
Hamiltonian at the initial moment $H(g(t=0))$, we do not have any
information about its eigenstates at other moments. Hence, we can
only quench the Hamiltonian back, so that it goes back to its
initial $H(g(t=0))$, and we can measure the final state (and compare
it with the initial state). The quench echo protocol (Eq. (\ref{1}))
ensures that the excitation probabilities in the forward quench
process and those in the backward quench process are similar (but
not identical; see Refs. \cite{bogdan06, Mosseri08}). As a result,
when the forward process is adiabatic (no excitations), so is the
backward quench process. Otherwise both the forward and the backward
processes are nonadiabatic, and the phase accumulated between
transitions (known as the St\"uckelberg phase) may result in
constructive or destructive interference \cite{sahel, nori}. Usually
the excitations in the forward and the backward processes cannot
cancel each other out (but see Ref. \cite{nori}).
Hence, through this quench echo method, without knowing about the
eigenenergies and eigenstates, one is able to evaluate the
adiabaticity of an arbitrary evolution in most cases. Nevertheless,
in some special cases (e. g., impulse evolution, which is so fast
that the state of the system is frozen) the final fidelity is equal
to unity, but the process may not be adiabatic. A solution to this
problem is to let the system evolve freely for some time before the
backward quench. We will discuss this in detail in the next section.
By utilizing quench echo one can even find a uniformly adiabatic
quench protocol for a given Hamiltonian by repeating the above
process with different quench time scales.

\section{A case study: Ising chain in a transverse magnetic field}

\begin{figure*}[ht]
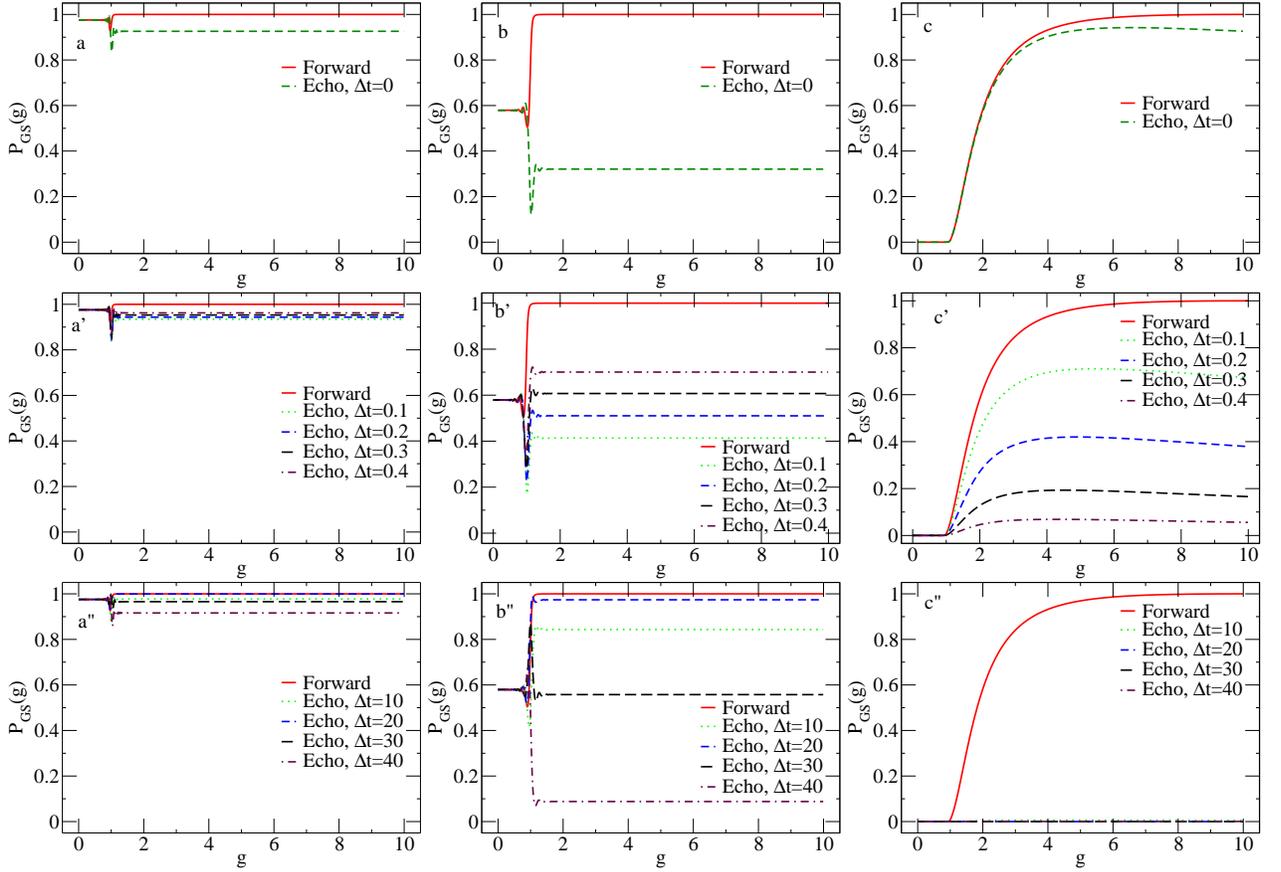

\begin{center}
\includegraphics[width=5.5cm, clip]{fig1a.eps}  \includegraphics[width=5.5cm, clip]{fig1b.eps}  \includegraphics[width=5.5cm, clip]{fig1c.eps}
\includegraphics[width=5.5cm, clip]{fig1d.eps}  \includegraphics[width=5.5cm, clip]{fig1e.eps}  \includegraphics[width=5.5cm, clip]{fig1f.eps}
\includegraphics[width=5.5cm, clip]{fig1g.eps}  \includegraphics[width=5.5cm, clip]{fig1h.eps}  \includegraphics[width=5.5cm, clip]{fig1i.eps}
\end{center}
\caption{Three regimes of the quench dynamics: (a) nearly adiabatic
regime, $\tau_{Q}=150$; (b) the intermediate regime $\tau_{Q}=35$;
and (c) nearly impulse regime $\tau_{Q}=0.004$. The horizontal axis
represents the parameter $g(t)$, which varies between 0 and 10, and
the vertical axis represents the probability of the system being in
the instantaneous ground state during the evolution. There is a
quantum phase transition at $g_c=1$. The  red solid line represents
the forward quench (from $g_0=10$ to $g_T=0$) and the green dashed
line represents the quench echo (from $g_T=0$ to $g_0=10$). In both
the nearly adiabatic and the nearly impulse regimes, the fidelity at
final moment is close to unity. In the first row (a)-(c), there is
no time delay at the turnaround point. In the second row
(a$'$)-(c$'$), the delay time at $g_T=0$ is $\Delta t=0.1$, $0.2$,
$0.3$, $0.4$. In the third row (a$''$)-(c$''$), the delay time at
$g_T=0$ is $\Delta t=10$, $20$, $30$, $40$. The number of the spins
in the Ising chain is $N=50$.} \label{fig1}
\end{figure*}

We now use a simple model to demonstrate our central
ideas: Consider the quench dynamics of an Ising model in an transverse
magnetic field \cite{sachdev}. The time-dependent Hamiltonian is
\begin{equation}
H(t)=-J \sum_{i=1}^{N} \left[ \sigma_{i}^{x} \sigma_{i+1}^{x}+g(t) \sigma_{i}^{z} \right],  \label{3}
\end{equation}
where $J$ indicates the energy scale; $ \sigma_{i}^{\alpha}$,
$\alpha=x, y, z$ is the Pauli matrix on the $i$ th lattice site; and
$g(t)$ is the reduced strength of the magnetic field, which varies
with time. It is known that for this model there is a finite energy
gap $\Delta=2J \frac{\pi}{N}$ at $g=\cos(\pi/N)$ when the size of
the system $N$ is finite. For simplicity, we consider a linear
quench protocol
\begin{equation}
g(t)= \left \{
\begin{array}{c}
\begin{split}
&g_{0} - \frac{t}{\tau_{Q}} , (0<t< (g_{0} -g_T) \tau_{Q} ),  \\
2 g_T -&g_{0} + \frac{t}{\tau_{Q}}, ( (g_{0} -g_T) \tau_{Q}  <t<
2(g_{0}- g_T) \tau_{Q}),
\end{split}
\end{array}
\right. \label{4}
\end{equation}
where $\tau_{Q}$ is the time scale of the quench. The larger the
$\tau_{Q}$, the slower the quench.  In the forward quench the
strength of the magnetic field is ramped from $g=g_{0}$ to $g=g_T$
continuously, and in the quench echo, it is ramped back from $g=g_T$
to $g=g_{0}$, where $g_T$ is the turnaround point. Initially the
system is prepared in the ground state of $H(g= g_{0})$. When one
quenches the system at different rates (by choosing different
$\tau_{Q}$), the fidelity (\ref{2}) will be different.

The Hamiltonian of the Ising model (\ref{3}) can be decoupled into $N$ independent fermionic
modes \cite{sachdev}.
\begin{equation}
H(t)= \sum_{k} \Lambda_{k}(g(t)) \left[\left | + (t) \right \rangle_{k}
\left \langle + (t) \right |_{k} -  \left | -(t)  \right \rangle_{k} \left
\langle - (t) \right |_{k}\right],
\end{equation}
where $\left | + (t) \right \rangle_{k}$ and $\left | - (t) \right
\rangle_{k}$ are the two instantaneous eigenstates of the $k$ mode.
Their corresponding eigenenergies are $\pm \Lambda_{k}(g(t))$, and $\Lambda (g(t))=J
\sqrt {g(t)^{2} -2g(t) \cos{k} +1}$. Here $k=(2s+1)\pi/N$,
$s=0,1,2,\cdots,N/2-1$ is the wave vector, and the number of spins $N$ is even.

We write the Schr\"odinger equation $ i \hbar
\frac{\partial}{\partial t} \left \vert \Phi (t) \right\rangle =
H(t) \left \vert \Phi (t) \right\rangle$ in the instantaneous
eigenbases $\left | +(t) \right \rangle_{k} $ and $\left | - (t)
\right \rangle_{k}$ of $H(t)$, where $ \left \vert \Phi (t)
\right\rangle = \prod_{k} \alpha_{k} (t) \left | + (t) \right
\rangle_{k} +\beta_{k} (t) \left | - (t) \right \rangle_{k}$. For
simplicity we choose $\hbar=1$ hereafter. For both the forward
quench ($0<t< (g_{0}-g_{T}) \tau_{Q} $) and the backward
($(g_{0}-g_{T}) \tau_{Q}  <t<2(g_{0}-g_{T}) \tau_{Q} $) process ,
the Schr\"odinger equation can be written as
\begin{equation}
\imath \frac{d}{dt}
\left[ \begin{array}{c}
\alpha_{k}(t)\\
\beta_{k}(t)
\end{array}
 \right]
 =  \left[ \begin{array}{c c}
 \begin{split}
2 \Lambda_{k} (g(t)),  \frac{-i J^{3} \sin {k}}{2 \Lambda_{k}^{2} (g(t))}\frac{dg(t)}{dt} \\
 \frac{i J^{3} \sin {k}}{2 \Lambda_{k}^{2} (g(t))} \frac{dg(t)}{dt}, -2 \Lambda_{k} (g(t))
 \end{split}
\end{array}
 \right] \left[ \begin{array}{c}
\alpha_{k}(t)\\
\beta_{k}(t)
\end{array}
 \right],
\label{5}
\end{equation}
where the initial condition for Eq. (\ref{5}) is
$\alpha_{k} (t=0)=0$, $\beta_{k} (t=0)=1$. The modulus square of the overlap between the final
state of Eq. (\ref{5}) and the instantaneous ground state at
$g=g_{0}$ gives the fidelity (\ref{2})
\begin{equation}
F=P_{GS}(2(g_{0}- g_T) \tau_{Q})=\prod_{k>0} \left\vert \beta_{k}
(2(g_{0}-g_T) \tau_{Q}) \right\vert^{2}.\label{8}
\end{equation}
In the following, we will focus on the solution of the Eq.
(\ref{8}). We will consider both the numerical and the analytical
results.

\subsection{Kibble-Zurek mechanism and three regimes}

Before the quantitative study of the fidelity and its relation with
the time scales of the quench, we describe the Kibble-Zurek
mechanism (KZM) \cite{kibble,zurek} of second-order phase
transitions, which provides a quantitative understanding of the
quench process.
The KZM describes e.g., the relation between the density of
topological defects, which are generated during quenching across a
phase transition, and the time scale of the quench (see Ref.
\cite{jacek09review} for a good review). The KZM was first
introduced in the classical phase transitions \cite{kibble,zurek},
and later generalized to quantum phase transitions \cite{zurek05}.
In our study, however, we will not focus on the density of
topological defects, but on the adiabaticity of the evolution of the
system.

A quantum phase transition is characterized by a vanishing
excitation gap $ \Delta(g(t)) \approx \Delta_0|g(t)-g_c|^{z\nu}$ and
a divergent correlation length $\xi \approx \xi_0/|g(t)-g_c|^\nu$,
where $z$ and $\nu$ are the critical exponents, and $\Delta_0$ and
$\xi_0$ are constants \cite{sachdev}. We define a dimensionless
distance from the critical point $g_{c}$ by
\begin{equation}
\epsilon(t)=\frac{g(t)-g_c}{g_c}.
\end{equation}
A generic $\epsilon(t)$ can be linearized near the critical point $\epsilon(t)=0$ as \cite{explain}:
\begin{equation}
\epsilon(t)\approx -\frac{t}{\tau_{Q}}.
\end{equation}
There are two interlinked time scales during a quench: the system
reaction time given by the inverse of the gap
$\tau(\epsilon(t))=1/\Delta_0|g(t)-g_c|^{z\nu}$ and the time scale
of transition given by
$|g(t)-g_c|^{z\nu}/\frac{d}{dt}|g(t)-g_c|^{z\nu}$. Away from the
critical point the reaction time is small in comparison with the
time scale of transition and the evolution is adiabatic. Near the
critical point, however, the opposite situation occurs and the
evolution is  approximately impulse (the state of the system is
frozen out). The boundary $\hat{t}$ between the two regions is
determined by the relation $\tau(\epsilon(t))= \epsilon/ \dot{
\epsilon}|_{\hat{t}}$, or
\begin{equation}
\frac{ 1}{|g(\hat{t})-g_c|^{z\nu}}  \sim \frac{ |g(\hat{t})-g_c|^{z\nu}}{\frac{d}{dt}|g(\hat{t})-g_c|^{z\nu}}.
\end{equation}
That is, $\hat{t} \sim (\frac{\tau_{Q}}{\hat{t}})^{z\nu}$, which
gives $\hat{t} \sim \tau_{Q}^{\frac{z\nu}{1+z\nu}}$
\cite{zurek,zurek05}. For the Ising model, we have $z=\nu=1$
resulting in $\hat{t}  \sim \tau_{Q}^{\frac{1}{2}}$~\cite{zurek05}.
According to KZM when $t \in (-\hat{t}, \hat{t})$, the system will
not evolve (the wavefunction will be frozen). Outside this time
interval the system will evolve approximately adiabatically.

For an infinitely large system, the energy gap is vanishingly small
at the critical point, and one can always find a $\hat{t}$.
According to the KZM this implies that, no matter how slow one
quenches the Hamiltonian in an infinite system, the evolution across
the critical point can never be adiabatic. For a finite-size system,
however, there is a finite
energy gap even at the critical point. 
When one quenches the system sufficiently slowly (large $\tau_{Q}$),
$\hat{t}$ approaches very near the critical point where -- for a
finite systems -- scalings no longer hold. As a consequence, the KZM
does not lead to simple scaling, as $\tau(\epsilon(t))= \epsilon/
\dot{ \epsilon}|_{\hat{t}}$ leads to a more difficult equation which
has to be solved to obtain $\hat t$. \cite{zurek2000}. Indeed -- in
accord with the adiabatic theorem -- the KZM predicts that when
$\tau_Q$ is larger than the inverse of the gap the transition will
remain adiabatic throughout. Thus, a finite energy gap allows an
adiabatic evolution across the critical point when the Hamiltonian
is driven sufficiently slowly. This is the {\it adiabatic quench
regime}. By contrast, when one quenches the Hamiltonian very fast
(small $\tau_{Q}$), there is a big $\hat{t}$ and there is an
approximately impulse regime for $t \in (-\hat{t}, \hat{t})$ when
the quench is essentially instantaneous. In this time interval the
system will approximately cease to evolve -- its wavefunction will
be frozen. This is the so-called {\it impulse regime}
\cite{zurek05,bogdan05}. When one chooses a time scale of quench
$\tau_{Q}$ between the above two limiting cases, the system will
evolve adiabatically when either $t<-\hat{t}$ or $t>\hat{t}$, and
will be frozen when $t\in (-\hat{t}, \hat{t})$. We call this regime
the {\it intermediate regime}.
We can summarize the quench behavior as follows: For a finite-size
system, when $\tau_{Q}$ is large enough, the evolution will be
adiabatic; When $\tau_{Q}$ is extremely small, the state of the
system will be frozen; When $\tau_{Q}$ is in between these two
limiting cases, the process is in the intermediate regime.

\subsection{Numerical and analytical results}

Having obtained the qualitative understanding of the quench dynamics
from the above KZM arguments, in the following we will study the
Ising model quantitatively, and compare the results with the
estimates obtained above. We consider a spin chain with a finite
size $N=50$, and start evolving it at $g_0=10$ and let it turn
around at $g_T=0$. There is a finite energy gap for this system at
the quantum critical point $g_{c}=1$. We choose three different
quench time scales $\tau_{Q}=150$, $\tau_{Q}=35$, and
$\tau_{Q}=0.004$ which correspond to the adiabatic, intermediate,
and impulse regimes. The system evolves under the time-dependent
Schr\"odinger equation. We plot the probability $P_{GS}(g)$ in the
instantaneous ground state as a function of the controlling
parameter $g$ during the quench process in Fig. 1a-1c.

\begin{figure*}[ht]
\begin{center}
\includegraphics[width=8.5cm, clip]{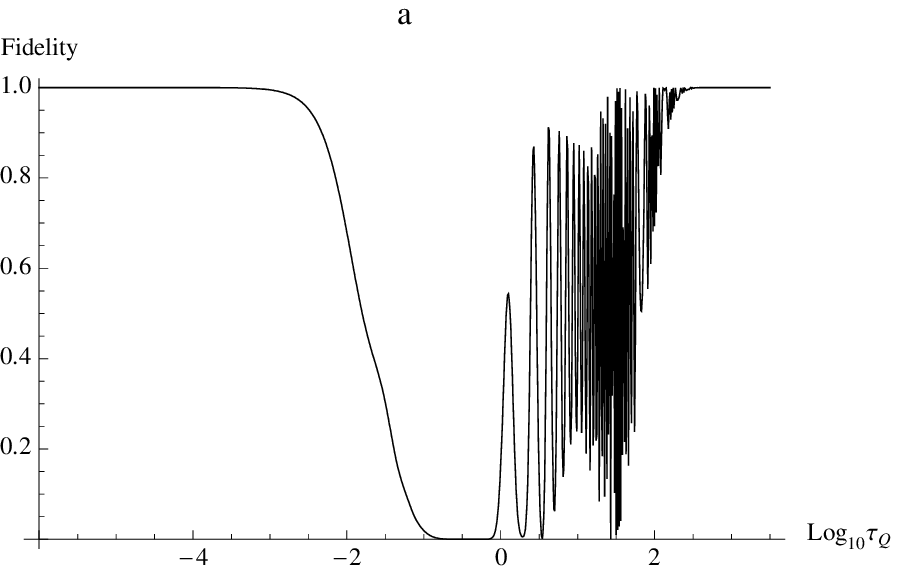}
\includegraphics[width=8.5cm, clip]{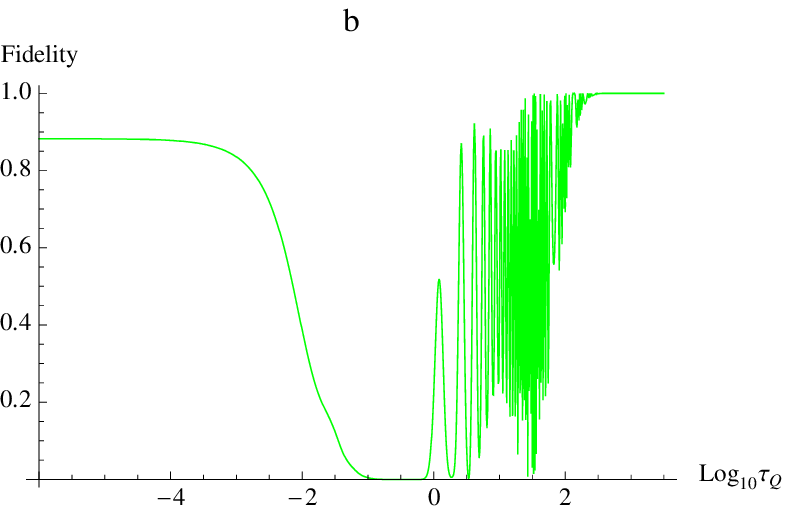}
\includegraphics[width=8.5cm, clip]{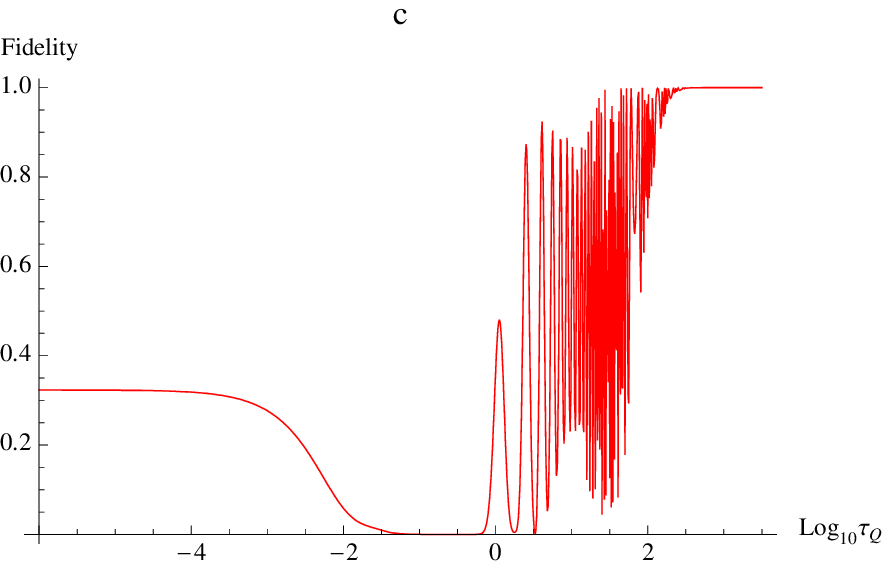}
\includegraphics[width=8.5cm, clip]{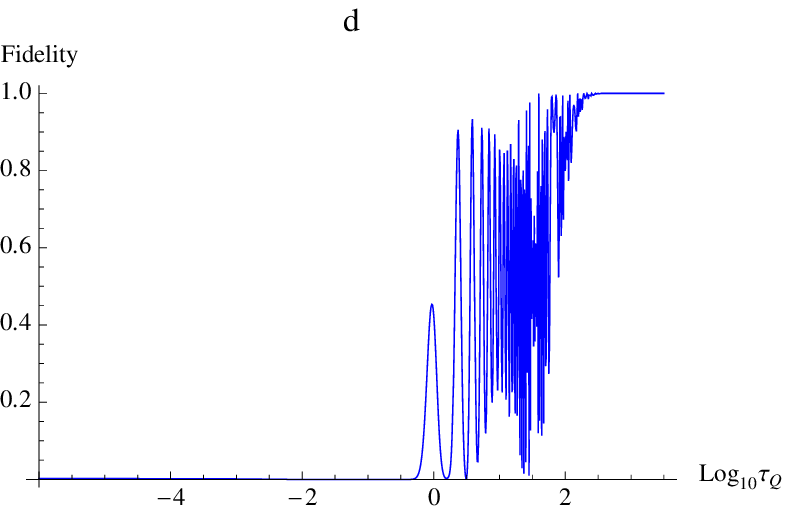}
\end{center}
\caption{The fidelity as a function of the time scale of quench
$\tau_{Q}$. Panels (a)-(d) represent different free evolution time
at turnaround point $g_T=0$ before the quench echo. Here $N=50$, and
start point at $g_0=10$, and the times of free evolutions are chosen
to be $\Delta t=0$, $\Delta t=0.1$, $\Delta t=0.3$, $\Delta t=0.7$.
It can be seen that in the adiabatic regime the fidelity is always
equal to unity, but in the impulse regime it is less than unity and
varies with the time of free evolution $\Delta t$.} \label{fig2}
\end{figure*}

From Fig. 1a it can be seen that when the time scale of the quench
is
relatively large, 
the system evolves almost adiabatically in the whole range of the
parameter $g_T=0<g<g_{0}=10$, $P_{GS}(g)$ is always close to unity
(except a tiny decay and partial revival at the critical points. So
is the fidelity of the quench echo (see Fig. 1a).

When the time scale of the quench is reduced to $\tau_{Q}=35$ (see Fig. 1b), the
quench dynamics enters the intermediate regime. It can be seen that away
from the critical point, the evolution is adiabatic. But near the
critical point,  the probability in the instantaneous ground state $P_{GS}(g)$
decays sharply and oscillates rapidly. This is
due to the interstate transitions at the anti-cross point. Soon
after passing through the quantum critical point the adiabatic evolution resumes.

When the time scale of the quench is further reduced to
$\tau_{Q}=0.004$ (almost instantaneous quench), the wave function of
the system is nearly frozen. Hence, the probability of being in the
instantaneous ground state is simply equal to the overlap of the
initial state and the instantaneous ground state. In the backward
quench, the same situation arises. Because in both the forward and
the backward quench, the wave functions of the system are frozen,
and hence are almost identical, the curves of $P_{GS}(g)$ of the
forward and the backward quench almost collapse onto the same curve
(see Fig. 1c), and the fidelity at $t=2(g_{0}-g_{T})\tau_{Q}$ is
close to unity.

We also plot the fidelity as a function of the quench time scale
$\tau_{Q}$ (see Fig. 2a). It can be seen that in both the impulse
regime $(\tau_{Q}< \thicksim10^{-3})$ and the adiabatic regime
$(\tau_{Q}>\thicksim10^{2.5})$, the fidelity is equal to unity. This
agrees with our intuition. Meanwhile, in the intermediate regime,
$\thicksim10^{-3}<\tau_{Q}<\thicksim10^{2.5}$, the fidelity
oscillates rapidly (see Fig. 2a).  When we plot the fidelity as a
function of the quench time $\tau_{Q}$, instead of $\ln{\tau_{Q}}$,
we found that there is a regular quasi-periodic oscillation (see
Fig. 3). We obtain an accurate expression of fidelity in the
intermediate regime,
\begin{equation}
\begin{split}
F\approx  \prod_{k>0}^{\pi/2} & \left |  e^{-2 \pi \tau_{Q}
\sin^{2}{k} }  e^{i \phi_k}  + \frac{2 \pi \tau_{Q} \sin^{2}{k} }{
\Gamma^{2}(1- i \tau_{Q} \sin^{2}{k})} e^{- \pi \tau_{Q} \sin^{2}{k}
}  e^{-i \phi_k} \right | ^{2},
\end{split}
\label{f}
\end{equation}
where $\phi_k= 2 \tau_{Q} [(-g_T + \cos{k})^{2} + \sin^{2}{k} \ln
\sqrt{4 \tau_{Q} (-g_T + \cos{k})^{2}}]$, and $ \Gamma(1- i \tau_{Q}
\sin^{2}{k})$ is the Gamma function (see Appendix A for details of
the derivation). From Fig. 3 it can be seen that the analytical
results agree with the numerical simulations, and that the fidelity
oscillates quasi-periodically with the increase of $\tau_{Q}$ as
expected. This oscillation can actually also be observed in Fig. 1
(see Fig. 1b' and Fig. 1c').

Numerical simulations agree with the results
obtained from the KZM very well, i.e., they account for three regimes that correspond to different $\tau_{Q}$.
We are especially interested in the first regime -- the adiabatic regime. From
Fig. 1a and Fig. 1c, it can be seen that in both the adiabatic regime and the
impulse regime, the fidelity is close to unity. In the next subsection,
we will introduce a method to eliminate the ``degeneracy" of the
adiabatic regime and the impulse regime.

\begin{figure}[ht]
\begin{center}
\includegraphics[width=8cm, clip]{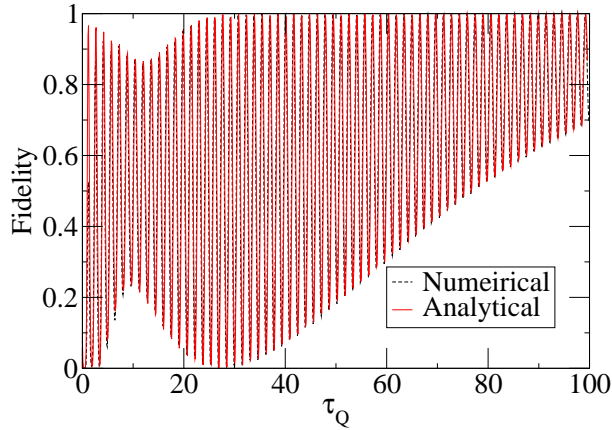}
\end{center}
\caption{Fidelity as a function of the time scale $\tau_Q$. Here the
black dashed line represents the numerical
 results whereas the red solid line represents the analytical results (Eq. (\ref{f})). It can be seen that the analytical results (Eq. (\ref{f}))
 agree well with the numerical results except for the case $\tau_Q\to 0$, where the quench process enters the impulse regime.
 The length of the spin chain is $N=50$, and the delay time at the turnaround point is $\Delta t=0$.}
\label{fig4}
\end{figure}

\subsection{Free evolution and decay of fidelity}

To distinguish the adiabatic and the impulse regime using quench
echo, one can let the system evolve freely for some time at the
turnaround point before quenching back. A study of the Landau-Zener
problem with waiting at the minimum gap has been reported in Ref.
\cite{sen10}. It was observed that the waiting influences the
excitation probability. Similarly in our study the free evolution at
the turnaround point leads to a decay in the fidelity in the impulse
regime, but makes no difference in the adiabatic regime (see Figs.
1a$'$-1c$'$ and Figs. 1a$''$-1c$''$). The reason is straightforward.
Let us first consider the adiabatic regime. Because the system is
always in its instantaneous ground state, the effect of the free
evolution is simply a global phase factor, which does not affect the
fidelity (see Figs. 1a$'$, 1a$''$, and Figs. 2b-2d). In the impulse
regime, the wavefunction before the free evolution is the ground
state of the initial Hamiltonian $H(g_{0}=10)$, and alternatively, a
superposition of the excited and the ground states of $H(g_T=0)$.
The excited and ground states acquire different phase factors during
the free evolution. Thus the wave function acquires relative phase
factors in its components and is no longer the ground state of
$H(g_T=0)$, but a superposition of its ground and excited states.
Hence, in the impulse regime when one quenches the system back to
the initial Hamiltonian $H(g_{0}=10)$, the system will no longer be
in its ground state, but in a superposition of the ground state and
the excited states. As a result, the fidelity is less than unity
(see Fig. 1c$'$, Fig. 1c$''$, and Figs. 2b-2d). The length of time
of the free evolution $\Delta t$ also influences the fidelity. One
can analytically calculate the fidelity as a function of the time of
free evolution $\Delta t$:
\begin{equation}
\left | - (g=+\infty) \right \rangle_{k} = \sin{\left[\frac{
\theta_{k}}{2} \right]} \left | + (g_{T}) \right \rangle_{k} +
\cos{\left[\frac{ \theta_{k}}{2}  \right]} \left | - (g_{T}) \right
\rangle_{k}, \nonumber
\end{equation}
where $\theta_{k}=\arctan(\frac{-\sin k}{\cos k -g_{T}})$. After
free evolution for $\Delta t$, the wave function becomes
\begin{equation}
\sin{\left[\frac{ \theta_{k}}{2}  \right]} e^ {-i \Lambda_{k}(g_{T})
\Delta t}\left | + (g_{T}) \right \rangle_{k} + \cos{\left[\frac{
\theta_{k}}{2}  \right]} e^ {i \Lambda_{k}(g_{T}) \Delta t}\left | -
(g_{T}) \right \rangle_{k}. \nonumber
\end{equation}
The fidelity can then be calculated as
\begin{equation}
F=\prod_{k>0} \left (1-\frac{\sin^{2} {k} \sin^{2}
{[\Lambda_k(g_{T}) \Delta t]}}{1-2 g_{T} \cos {k} +g^{2}_{T}}
\right).
\end{equation}
Note that for a fixed chain size, the value of $\Delta t$ needed to
scramble all the relevant phases is relevant to the range of the
spectrum of the system or the size of gap $\Lambda_{k}(g_{T})$ of
different $k$ at the turnaround point $g_{T}$. When
$\Lambda_{k}(g_{T})$ is very small, i.e., the energy spectrum of the
system is concentrated within a very small energy range, one needs
to wait for a long time in order to scramble all the relevant
phases: $\Delta t$ is inversely proportional to the energy scale $J$
of $\Lambda_{k}(g_{T})$. For a spin chain of $N=50$, when the time
of free evolution is very short, e.g., $\Delta t=0.1$, there is a
pronounced decay in the fidelity in the impulse regime (see Fig.
2b). The analytical result gives $F \approx 0.882$, which agrees
with the numerical result. The fidelity decreases with the increase
of time of the free evolution. The fidelity decays to $0.002$ when
$\Delta t = 0.7$ (see Fig. 2d).
Hence the quench echo with a free evolution at the turnaround point
can distinguish the adiabatic and the
impulse regime. Our numerical results confirm our theoretical predictions.



\section{Beyond the linear quench}

In the above discussion, we focused on the linear
quench. One may repeat the above process with different $\tau_{Q}$ until one finds the smallest
$\tau_{Q}^{c}$, under which the process is sufficiently adiabatic,
for example $F \ge 0.9$. Nevertheless, the linear quench with
$\tau_{Q}^{c}$ obtained above may waste a lot of time. The reason is
obvious: in different regions of the parameter $g$, the energy gaps
are different. According to KZM, different energy gaps correspond to
different relaxation time $\tau$. For a linear quench protocol, we
are treating the whole range of the parameter uniformly, and the
relaxation time is determined by the global minimal energy gap.
Thus, we waste a lot of time. Usually we want to ensure that the
process not only nearly adiabatic but also as fast as possible. In
the following we will consider nonlinear quench.

\subsection{Adjusting quench rate to the instantaneous gap}

An improved scheme is to divide the whole range of the parameter
into many, e.g., $M$, parts with equal length $(g_{0}-g_T)/M$, and
then apply the above linear quench protocol to these ranges
separately to find the uniformly adiabatic quench for each range
$\tau_{Q}^{ci}$, $i=1,2, \cdots, M$.
We can also use the KZM to find a uniformly adiabatic quench
protocol. From the discussion in Section III.A we know that the
transition time scale is given by the absolute value of
$\Delta(g(t))/\frac{d}{dt}\Delta(g(t))$. Meanwhile, the relaxation
time scale is given by $1/\Delta(g(t))$. When the former is many
times larger than the latter, the process should be uniformly
adiabatic. That is, when the parameter $g(t)$ satisfies the relation
\begin{equation}
\left | \frac{ \Delta(g(t))}{\frac{d}{dt}\Delta(g(t))} \right |= \frac{\gamma }{\Delta(g(t))}. \label{11}
\end{equation}
where $\gamma$ is a constant many times larger than unity, e.g.,
$\gamma=10$, the process is uniformly adiabatic in the sense that
the ratio of two time scales remains a constant. Such a quench
scheme is better than the linear quench. The solution to the above
ordinary differential equation is
\begin{equation}
\Delta(g(t))=\frac{1}{\mp\frac{1}{\gamma }t+c},  \label{12}
\end{equation}
where $\mp$ corresponds to the sign of $\Delta/\dot \Delta$ on the
left-hand-side of Eq. (\ref{11}) being positive or negative, and $c$
is a constant of integration. For simplicity $c$ can be chosen such
that at $t=0$ $\Delta$ in Eq. (\ref{12}) is the minimal gap.
Now, we know exactly the energy gap as a function of the controlling
parameter (see Fig. 4a) \cite{gap}
\begin{equation}
\begin{split}
\Delta(g(t))= &2J \sqrt{1-2g(t) \cos (\frac{\pi}{N})+g^{2}(t)}. \label{13}
\end{split}
\end{equation}
Therefore, $c$ can be determined by
$g(t=0)=\cos{(\pi/N)}$. Combining Eqs. (\ref{12}) and (\ref{13}), we
find the following uniformly adiabatic quench protocol (see Fig. 4b)
\begin{equation}
g_{\mathrm{KZ}}(t)= \left \{
\begin{array}{c}
\begin{split}
& \cos (\frac{\pi}{N})- \sqrt{-\left( \sin (\frac{\pi}{N})\right)^{2}+ \frac{(\gamma )^2}{4J^{2} \left( t+  \frac{\gamma }{2J \sin{(\pi/N)}} \right)^{2}} },  (-\frac{\gamma }{2J \sin{(\pi/N)}} <t< 0)  \\
& \cos (\frac{\pi}{N})+\sqrt{-\left( \sin
(\frac{\pi}{N})\right)^{2}+ \frac{(\gamma  )^2}{4J^{2} \left(
-t+ \frac{\gamma }{2J \sin{(\pi/N)}} \right)^{2}} }, (0  <t<
\frac{\gamma }{2J \sin{(\pi/N)}})
\end{split}
\end{array}.
\right. \label{14}
\end{equation}
It can be seen that the time required for the whole process
(quenching the controlling parameter from $g=0$ to $g=\infty$) is
given by
\begin{equation}
\Delta t_{\mathrm{KZ}}=\frac{\gamma }{J \sin{(\pi/N)}} \approx \frac{\gamma  N}{J \pi}, \label {15}
\end{equation}
or $\Delta t_{\mathrm{KZ}}= \frac{2\gamma }{\Delta_{\min}}$, which
is proportional to the chain size $N$ and the ratio $\gamma$, and
inversely proportional to the minimum energy gap $\Delta_{\min}=2J
\sin \frac{\pi}{N}\approx 2J \frac{\pi}{N}$. In linear quench the
minimal time required for the adiabatic evolution grows with the
system size like $N^{2}$ \cite{jacek}. The quench scheme of Eq.
(\ref{14}) is obviously better. This agrees with previous studies
that ``non-linear" quench can improve the adiabaticity (minimize
excitation) \cite{jacek09review, nonlinear1, nonlinear2}. The energy
gap and the protocols for uniformly adiabatic quench (Eq.
(\ref{14})) are shown in Fig. 4.
\begin{figure}[ht]
\begin{center}
\includegraphics[width=8cm, clip]{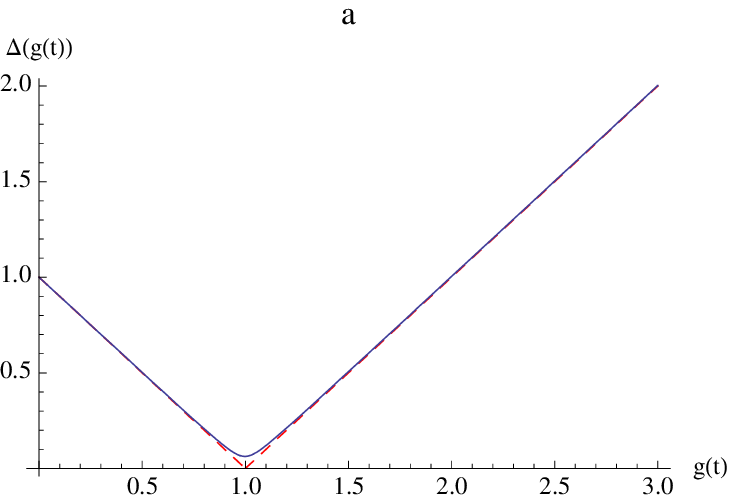}
\includegraphics[width=8cm, clip]{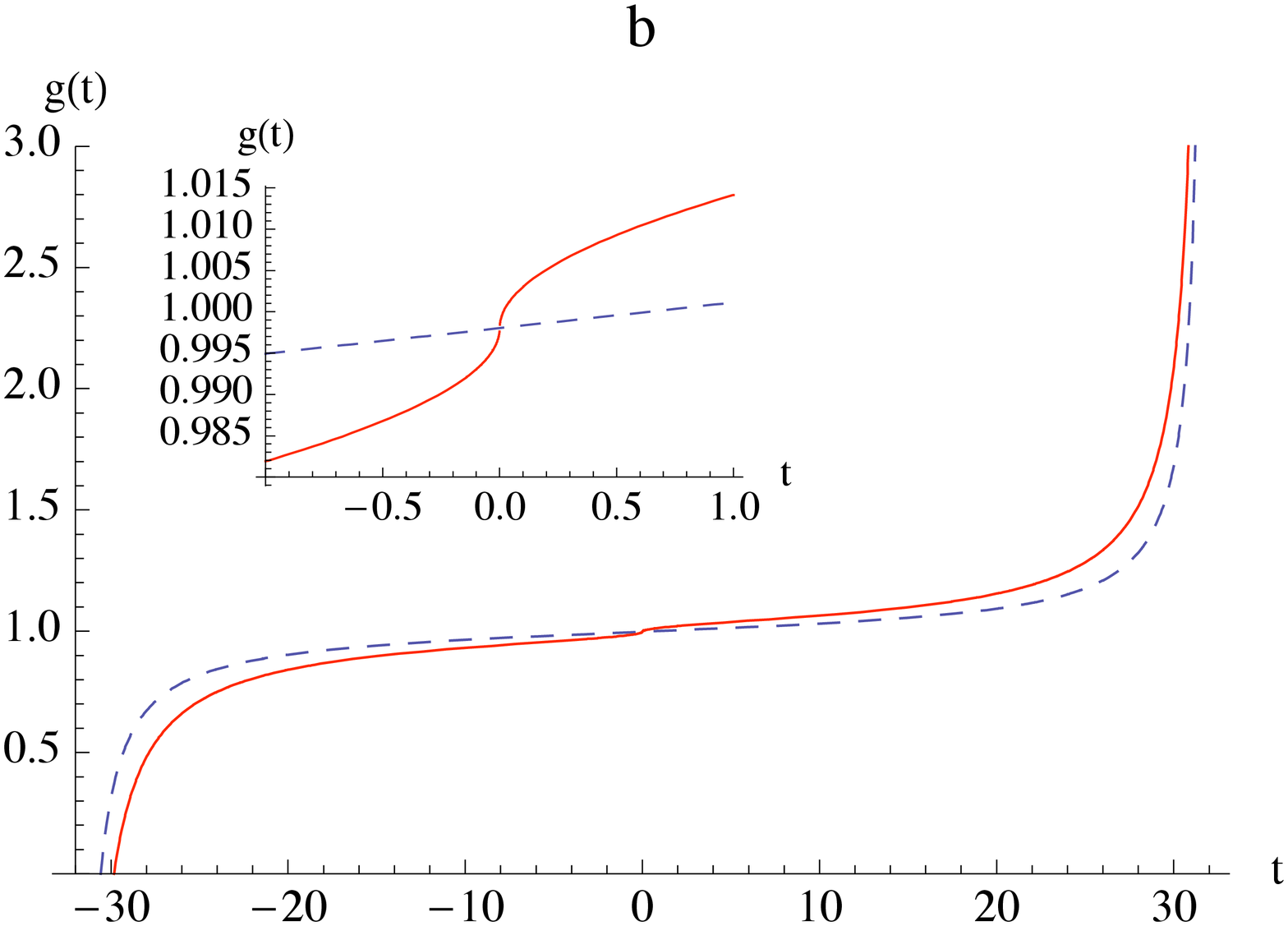}
\end{center}
\caption{ (a) Energy gap as a function of the controlling parameter
$g$ for a finite size chain. Here the solid line indicates the spin
chain of $N=50$ and the dashed line indicates the gap of an infinite
chain. (b) The uniformly adiabatic quench protocol associated with
the KZM criterion (solid) (\ref{14}) and associated with RC
criterion (dashed) (\ref{17}). Here the ratio $\gamma=2$, and the
energy scale $J=1/2$. We have chosen the condition
$g(t=0)=\cos{(\pi/N)}$.} \label{fig4}
\end{figure}

Note that the criterion for uniformly adiabatic evolution derived
from the KZM (Eq. (\ref{11})) is similar, but not identical, to the
criterion proposed by Roland and Cerf that was derived from the
quantum adiabatic theorem (See Eq. (17) of reference \cite{cerf}).
In the RC model the energy gap is inversely proportional to $\sqrt
{N}$, and the minimum time required is proportional to $\sqrt {N}$.
But in the Ising chain, the energy gap is inversely proportional to
$N$, and the minimum time required is proportional to $N$. It can be
proved that if one uses Roland and Cerf's criterion to evaluate the
minimum time required for the uniformly adiabatic evolution, the
minimum time is also proportional to $N$.

It is interesting to compare the two criteria for uniformly
adiabatic evolution in the Ising chain. In the following we will
first solve the equation of the quench protocol for uniformly
adiabatic evolution $g_{\mathrm{RC}}(t)$ associated with the Roland
and Cerf's criterion and then simulate the dynamic evolution of the
Ising chain with both $g_{\mathrm{RC}}(t)$ and $g_{\mathrm{KZ}}(t)$.
We will fix the time of quench process, and compare the fidelity of
the two protocols. The Roland and Cerf's criterion (see Eq. (17) of
Ref. \cite{cerf}) is
\begin{equation}
\left | \frac{d}{dt} \left| g(t)-g_{c}\right| \right |= \frac{1}{\gamma ' } \Delta^{2}(g(t)), \label{16}
\end{equation}
where $\gamma '$ is the ratio between the two time scale. Obviously,
when $N\to \infty$, $\Delta(g)=\left| g(t)-g_{c}\right|$ is valid
for arbitrary $g$. In this respect, the two criteria, Eq. (\ref{11})
and Eq. (\ref{16}), are equivalent. Nevertheless, when $N$ is
finite, the two criteria differ slightly because the gap $\Delta(g)$
deviates from $\left| g(t)-g_{c}\right|$ near the critical point
(see Eq. (\ref{13}) and Fig. 4a). As a result there is a small
discrepancy in the quench protocols $g_{\mathrm{RC}}(t)$ and
$g_{\mathrm{KZ}}(t)$ associated with two criteria, especially when
$g(t)$ is close to $g_{c}$.

By substituting Eq. (\ref{13}) into Eq. (\ref{16}), we obtain the
quench protocol:
\begin{equation}
g_{\mathrm{RC}}(t)=  \cos (\frac{\pi}{N}) + \sin (\frac{\pi}{N})
\tan{\left( \frac{2J \sin (\frac{\pi}{N})}{  \gamma '}t\right)},
(-\frac{ \gamma ' N}{4J}<t<\frac{ \gamma ' N}{4J}).
\label{17}
\end{equation}
We plot the solution $g_{\mathrm{RC}}(t)$ along with
$g_{\mathrm{KZ}}(t)$ in Fig. 4b. There is a ``kink'' at the
anti-cross point of the energy levels in $g_{\mathrm{KZ}}(t)$
associated with the KZM criterion, but there is none in
$g_{\mathrm{RC}}(t)$ associated with the RC criterion (see Fig. 4b
and the inset). Although there is a singularityÓ (divergent time
derivative of g) in $g_{\mathrm{KZ}}(t)$ at $t=0$, the time interval
of this region is vanishingly small. As a result the total change in
$g$ in this singular region is very small (See Eq. 4b), and the
eigenstates of H(t) do not change significantly within it. This is
in the same spirit as quantum fidelity \cite{zanardi06}, where
fidelity susceptibility diverges at quantum critical point, but the
fidelity is nonzero indicating that the ground state does not
changes significantly. Hence, the 'kink' at $t=0$ will not lead to a
lot of excitations. Our simulation verifies this point. Similar to
Eq. (\ref{15}), we obtain the time required for the uniformly
adiabatic evolution
\begin{equation}
\Delta t_{\mathrm{RC}}=\frac{\gamma ' N}{2J}. \label{18}
\end{equation}
Comparing Eq. (\ref{15}) and Eq. (\ref{18}), we find that when
$\gamma '=\frac{2}{\pi}\gamma$, the time required for two criteria
are equal. In the following we will simulate the dynamics of the
Ising chain under the two quench protocols: Eq. (\ref{14}) and Eq.
(\ref{17}). Substituting Eq. (\ref{14}) and Eq. (\ref{17}) into Eq.
(\ref{5}), one obtains the instantaneous fidelity as a function of
the time $F= \left | \beta(t) \right| ^{2}$ associated with two
criteria. We plot the fidelity as a function of the time in Fig. 5.
When one chooses a different initial condition, the fidelity as a
function of the time differs a lot. In the left panel of Fig. 5, we
plot the fidelity as a function of the time quenching from
$t=-\frac{\gamma}{2J \sin{(\pi/N)}}$ ($g=-\infty$). The fidelity
associated with the Roland and Cerf criterion decays when the system
is near the anti-crossing point, and then revives. But the fidelity
associated with the KZM does not change much, and remains close to
unity all the time. At $t=\frac{\gamma  }{2J \sin{(\pi/N)}}$
($g=\infty$), the fidelity associated with the Roland and Cerf's
criterion is a bit higher than that associated with the KZM. In the
right panel of Fig. 5, we plot the fidelity as a function of the
time starting from $t=- \frac{1}{3}\frac{\gamma }{2J
\sin{(\pi/N)}}$. In this case the fidelity associated with the
Roland and Cerf's criterion oscillates rapidly and finally reaches a
stable value around $0.85$. By contrast the fidelity associated with
the the KZM does not oscillate and remains very close to unity. From
the above facts, we conclude that in some cases, the Roland and
Cerf's criterion is better than the KZM criterion, but in some other
cases, it is worse. Hence, we cannot say which criterion is
definitely better, but the KZM provides new insights into the
conditions for uniformly adiabatic evolution.

One is tempted at this point to undertake a variational study in
search of optimal quenches. While such a study is beyond the scope
of this paper, we note that in practical applications (e.g.,
adiabatic quantum computing) optimization would involve not just
varying rate, but (as it was done in Fig. 5) also the starting and
final points of the quenches can be brought closer to the ``critical
point''. Resulting errors can be detected and the correct result can
be ascertained by repeating the computation many times.
\begin{figure}[ht]
\begin{center}
\includegraphics[width=8cm, clip]{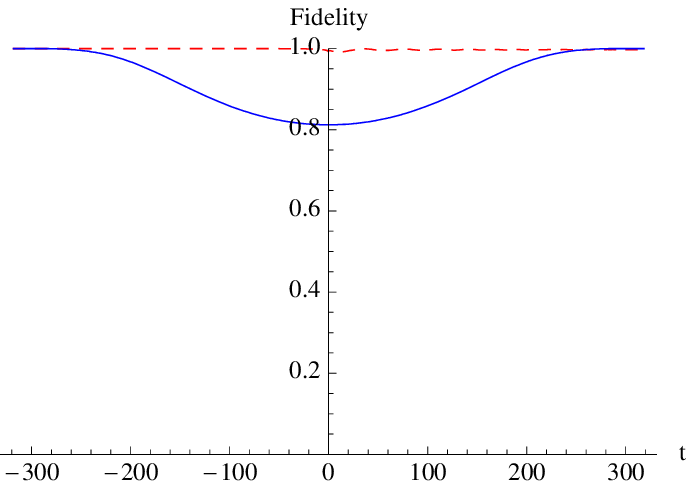}
\includegraphics[width=8cm, clip]{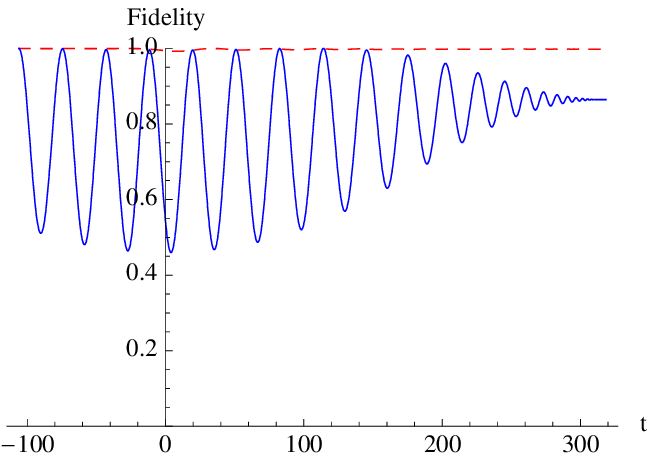}
\end{center}
\caption{ Comparison of the Roland and Cerf criterion and the KZM
criterion. The horizontal axis indicates the time, and the vertical
axis depicts the instantaneous fidelity. The solid (dashed) line
describes the fidelity as a function of the time $t$ associated with
the Roland and Cerf (KZM) criterion. Left panel: The evolution
starts from the ground state of $t=-\frac{\gamma N}{2J \pi}$
($g=-\infty$). Right panel: The evolution starts from the ground
state of $t=- \frac{1}{3}\frac{\gamma N}{2J \pi}$.} \label{fig5}
\end{figure}

\subsection{Gauging the distance from the adiabatic quench}

Fig. 2 indicates that when the time scale of the quench $\tau_{Q}$
is in the range $\tau_{Q} \in (10^{-1}, 1)$, the fidelity is almost
equal to zero. However, this does not reveal how far the quench is
from the adiabaticity. For example when one out of many ($N=50$ in
our numerical simulation) modes get excited, the fidelity will decay
to nearly zero due to the orthogonality of one mode. But, in a
sense, the system is still close to the ground state, as all but one
excited state are empty. In this sense, the fidelity is not a good
criterion for measuring how far away the quench is from the
adiabatic evolution.

A better gauge of the distance of a quench process from the
adiabaticity may be obtained using 
other variables, such as the magnetization per site along the
direction of the external magnetic field
\begin{equation}
m=\frac{1}{N} \sum_{i=1}^{N}{\left\langle
\sigma_{i}^{z}\right\rangle} = \frac{1}{N}\sum_{k=1}^{N} 2
\left\vert \beta_{k} ((g_{0}-2 g_T) \tau_{Q}) \right\vert^{2} - 1.
\end{equation}
When the magnetic field is large, the ground state corresponds to
$m=1$. We plot the final magnetization as a function of $\tau_{Q}$
in FIg. 6. In the range of $\tau_{Q}\in (10^{-1}, 1)$, the fidelity
is vanishingly small, but the magnetization per site is still large.
This indicates that the system is not very far away from the
instantaneous ground state. Moreover, when one delays for some time
at the turnaround point, the magnetization per site of the impulse
regime will decrease, but that of the adiabatic regime will not (see
Fig. 6). This is similar to the fidelity and agrees with our
intuition. Last but not least, the magnetization is experimentally
easier to the measure than the fidelity, and it has been used as a
tool to study the adiabaticity of quantum dynamics in Ref.
\cite{anarb}.

One can also use the kink density
$\frac{1}{2}\sum_{i}(1-\left\langle \sigma_{i} \sigma_{i+1}
\right\rangle)$ \cite{jacek} as a measure of the distance of the
system from the adiabaticity. The relation between the fidelity and
the density of defects has been studied in Ref. \cite{jacek09prb}.
Other variables, such as the residual energy \cite{fazio}, can be
also used to gauge the distance from the adiabatic quench. Such
obvious measures of how far the quench is from the adiabaticity work
well in the one-dimensional Ising model, but finding their useful
analogues in other situations (e.g., adiabatic quantum computing)
may not be easy.

\begin{figure}[ht]
\begin{center}
\includegraphics[width=8cm, clip]{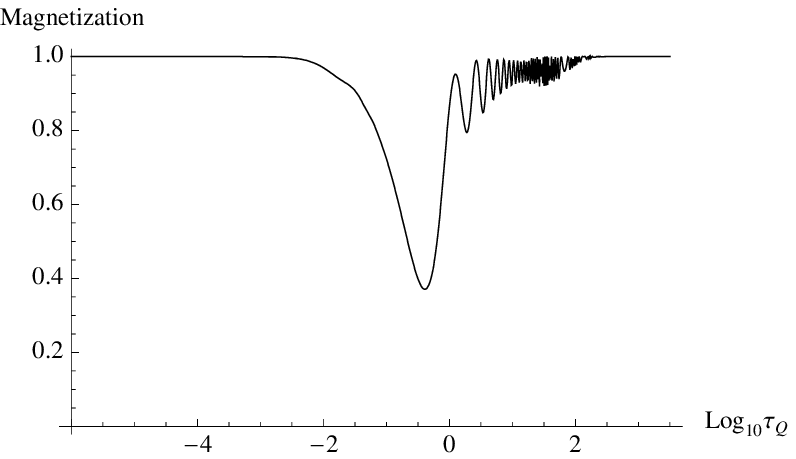}
\includegraphics[width=8cm, clip]{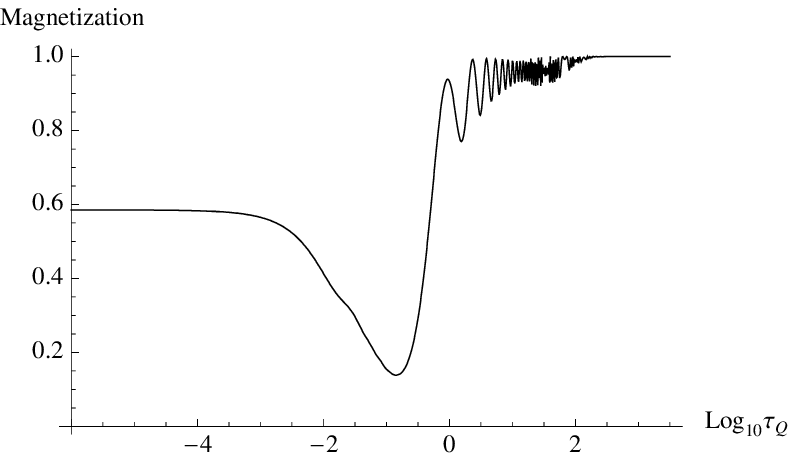}
\end{center}
\caption{ Final magnetization of the Ising chain as a function of
the time scale of the quench $\tau_{Q}$. All the parameters are the
same as those in Fig. 2(a). One can see that the spin chain is not
very far away from equilibrium except when the time scale of the
quench is in the range $\tau_{Q}\in(10^{-1},1)$. Left panel: delay
time $\Delta t=0$; Right panel: delay time $\Delta t=0.7$.}
\label{fig5}
\end{figure}


\section{Summary and conclusion}
We have proposed a strategy to test the adiabaticity without knowing
either the eigenstates or eigenenergies of the Hamiltonian. Instead
of having to find the gap of the Hamiltonian, and then using the
quantum adiabatic approximation to evaluate the adiabaticity, one
can use a quench echo to evaluate the adiabaticity of an evolution.
The underlying mechanism is that when the time scale of the quench
is large in comparison with the inverse of the energy gap, both the
forward and the backward evolutions are adiabatic. As a result, the
fidelity of the initial state and the final state is close to unity.
Otherwise, the evolution is not adiabatic, and the fidelity is less
than unity. The method for testing the adiabaticity of an evolution
presented in this paper is universally valid. It does not depend the
model or the validity of conditions for adiabatic approximation. We
further proposed a method for finding the uniformly adiabatic quench
protocol based on the KZM, and discussed the problem of gauging how
non-adiabatic is a quench. Given the importance of the adiabaticity
in various applications, we believe that our results will be broadly
applicable, and may be useful in experimental applications.



\acknowledgements This work is supported by U.S. Department of
Energy through the LANL/LDRD Program. We thank Bogdan Damski, Anarb Das, and
Rishi Sharma for helpful discussions.

\begin{appendix}
\setcounter{section}{0} \setcounter{equation}{0}
\renewcommand{\thesection}{\Alph{section}}

\section{fidelity in the intermediate regime}
From Refs. \cite{jacek,bogdan}, we know that in the wave vector $k$
representation, the Schr\"odinger equation for the forward quench
can be rewritten as Landau-Zener type equations (see Eq. (\ref{5})
for a comparison):
\begin{equation}
i\frac{d}{dt'}
\left[
\begin{array}{c}
v_{k} (t') \\
u_{k} (t')
\end{array}
\right]
=
\frac{1}{2}\left[
\begin{array}{lr}
\frac{t'}{\tau_Q'} & 1 \\
1  & -\frac{t'}{\tau_Q'}
\end{array}
\right]
\left[
\begin{array}{c}
v_{k} (t')\\
u_{k} (t')
\end{array}
\right],
\label{LZ}
\end{equation}
where $t'=4\tau_Q \sin k [-g(t)+\cos k]$ and $\tau_Q'=4\tau_Q
\sin^{2} k$. This equation can be solved in terms of Weber
functions. The initial conditions are $v_k(t'=-\infty)=0$ and
$u_k(t'=-\infty)=1$. The solution for this equation is \cite{bogdan}
\begin{equation}
\begin{split}
u_{k}(t) \approx &e^{- \pi \tau_{Q} \sin^{2}{k} }  e^{i  \tau_{Q} [(-g + \cos{k})^{2} + \sin^{2}{k} \ln \sqrt{4 \tau_{Q} (-g + \cos{k})^{2}}]},\\
v_{k}(t) \approx &\frac{ \sqrt {2 \pi \tau_{Q} \sin^{2}{k} }}{ \Gamma (1- i \tau_{Q} \sin^{2}{k})} e^{- \frac{\pi \tau_{Q} \sin^{2}{k}}{2} } e^{-i  \tau_{Q} [(-g + \cos{k})^{2} + \sin^{2}{k} \ln \sqrt{4 \tau_{Q} (-g + \cos{k})^{2}}]}.
\end{split}
\label{a2}
\end{equation}
Similarly, we can obtain the solution for the quench echo. Combining
the forward and the quench echo process, we find the solution of the
fidelity (\ref{f}). It is worth pointing out that the above solution
is only good for the turnaround point far away from the critical
point $ g_{T} \ll 1$. For example $g_{T}=0.5$.

\end{appendix}


\begin{thebibliography}{99}
\bibitem{clarification} In recent years the conditions for the adiabatic
approximation have become a subject of controversy
\cite{controversy,anarb}. The standard criterion is found to be
neither sufficient nor necessary \cite{jianda}. Additional
assumptions are required to ensure the stated conditions to be
rigorous. See Ref. \cite{nori} for a clarification of weakness of
adiabaticity conditions.
\bibitem{footnote} Reference \cite{sen08} studied the density of topological defects
under such quench echo. But our emphasis is different.
\bibitem{explain} Kibble-Zurek mechanism with nonlinear quench schemes is also discussed recently, see Refs. \cite{jacek09review,nonlinear1,nonlinear2}.
\bibitem{gap} The energy gap of other modes is much larger than the considered mode $k=\pi/N$. So one usually consider only the excitations (if there is any) in the smallest mode $k=\pi/N$, and ignore the excitations in the other modes
\end{thebibliography}
\end{document}